%% file: temp_body.tex
\documentclass[]{pasj01}
\draft

\begin{document} 
\Received{}
\Accepted{}

\title{A Substellar Companion to Pleiades~HII~3441}

\author{Mihoko \textsc{Konishi}\altaffilmark{1,$\ast$}}
 \email{mihoko.konishi@nao.ac.jp}
\author{Taro \textsc{Matsuo},\altaffilmark{2,$\ast$}}
 \email{matsuo@ess.sci.osaka-u.ac.jp}
\author{Kodai \textsc{Yamamoto}\altaffilmark{3}}
\author{Matthias \textsc{Samland}\altaffilmark{4}}
\author{Jun \textsc{Sudo}\altaffilmark{2}}
\author{Hiroshi \textsc{Shibai}\altaffilmark{2}}
\author{Yoichi \textsc{Itoh}\altaffilmark{5}}
\author{Misato \textsc{Fukagawa}\altaffilmark{6}}
\author{Takahiro \textsc{Sumi}\altaffilmark{2}}
\author{Tomoyuki \textsc{Kudo}\altaffilmark{7}}
\author{Jun \textsc{Hashimoto}\altaffilmark{8}}
\author{Masayuki \textsc{Kuzuhara}\altaffilmark{8, 9}}
\author{Nobuhiko \textsc{Kusakabe}\altaffilmark{8}}
\author{Lyu \textsc{Abe}\altaffilmark{10}}
\author{Eiji \textsc{Akiyama}\altaffilmark{1}}
\author{Wolfgang \textsc{Brandner}\altaffilmark{4}}
\author{Timothy D. \textsc{Brandt}\altaffilmark{11}}
\author{Joseph C. \textsc{Carson}\altaffilmark{12}}
\author{Markus \textsc{Feldt}\altaffilmark{4}}
\author{Miwa \textsc{Goto}\altaffilmark{13}}
\author{Carol A. \textsc{Grady}\altaffilmark{14, 15}}
\author{Olivier \textsc{Guyon}\altaffilmark{7, 16, 8}}
\author{Yutaka \textsc{Hayano}\altaffilmark{1}}
\author{Masahiko \textsc{Hayashi}\altaffilmark{1}}
\author{Saeko S. \textsc{Hayashi}\altaffilmark{7}}
\author{Thomas \textsc{Henning}\altaffilmark{4}}
\author{Klaus W. \textsc{Hodapp}\altaffilmark{17}}
\author{Miki \textsc{Ishii}\altaffilmark{1}}
\author{Masanori \textsc{Iye}\altaffilmark{1}}
\author{Markus \textsc{Janson}\altaffilmark{18}}
\author{Ryo \textsc{Kandori}\altaffilmark{1}}
\author{Gillian R. \textsc{Knapp}\altaffilmark{19}}
\author{Jungmi \textsc{Kwon}\altaffilmark{20, 21}}
\author{Michael W. \textsc{McElwain}\altaffilmark{15}}
\author{Kyle \textsc{Mede}\altaffilmark{20}}
\author{Shoken \textsc{Miyama}\altaffilmark{22}}
\author{Jun-Ichi \textsc{Morino}\altaffilmark{1}}
\author{Amaya \textsc{Moro-Mart\'{i}n}\altaffilmark{23}}
\author{Tetsuo \textsc{Nishimura}\altaffilmark{7}}
\author{Daehyeon \textsc{Oh}\altaffilmark{24, 1, 25}}
\author{Tae-Soo \textsc{Pyo}\altaffilmark{7}}
\author{Eugene \textsc{Serabyn}\altaffilmark{26}}
\author{Joshua E. \textsc{Schlieder}\altaffilmark{27, 4}}
\author{Takuya \textsc{Suenaga}\altaffilmark{1, 25}}
\author{Hiroshi \textsc{Suto}\altaffilmark{1, 8}}
\author{Ryuji \textsc{Suzuki}\altaffilmark{1}}
\author{Yasuhiro H. \textsc{Takahashi}\altaffilmark{20, 1}}
\author{Michihiro \textsc{Takami}\altaffilmark{28}}
\author{Naruhisa \textsc{Takato}\altaffilmark{7}}
\author{Hiroshi \textsc{Terada}\altaffilmark{1}}
\author{Christian \textsc{Thalmann}\altaffilmark{29}}
\author{Edwin L. \textsc{Turner}\altaffilmark{19, 30}}
\author{Makoto \textsc{Watanabe}\altaffilmark{31}}
\author{John P. \textsc{Wisniewski}\altaffilmark{32}}
\author{Toru \textsc{Yamada}\altaffilmark{21}}
\author{Hideki \textsc{Takami}\altaffilmark{1}}
\author{Tomonori \textsc{Usuda}\altaffilmark{1}}
\author{Motohide \textsc{Tamura}\altaffilmark{20, 1, 8}}

\altaffiltext{1}{National Astronomical Observatory of Japan, Tokyo, Japan}
\altaffiltext{2}{Department of Earth and Space Science, Graduate School of Science, Osaka University, Osaka, Japan}
\altaffiltext{3}{Department of Astronomy, Faculty of Science, Kyoto University, Kyoto, Japan}
\altaffiltext{4}{Max Planck Institute for Astronomy, Heidelberg, Germany}
\altaffiltext{5}{Nishi-Harima Astronomical Observatory, Hyogo, Japan}
\altaffiltext{6}{Division of Particle and Astrophysical Science, Graduate School of Science, Nagoya University, Nagoya, Japan}
\altaffiltext{7}{Subaru Telescope, National Astronomical Observatory of Japan, HI, USA}
\altaffiltext{8}{Astrobiology Center, Tokyo, Japan}
\altaffiltext{9}{Department of Earth and Planetary Sciences, Tokyo Institute of Technology, Tokyo, Japan}
\altaffiltext{10}{Laboratoire Lagrange, Universit\'{e} de Nice-Sophia Antipolis, Nice, France}
\altaffiltext{11}{Astrophysics Department, Institute for Advanced Study, NJ, USA}
\altaffiltext{12}{Department of Physics and Astronomy, College of Charleston, SC, USA}
\altaffiltext{13}{Universit\"ats-Sternwarte M\"unchen, Ludwig-Maximilians-Universit\"at, M\"unchen, Germany}
\altaffiltext{14}{Eureka Scientific, CA, USA}
\altaffiltext{15}{Exoplanets and Stellar Astrophysics Laboratory, code 667, Goddard Space Flight Center, MD, USA}
\altaffiltext{16}{The University of Arizona, AZ, USA}
\altaffiltext{17}{Institute for Astronomy, University of Hawaii, HI, USA}
\altaffiltext{18}{Department of Astronomy, University of Stockholm, Stockholm, Sweden}
\altaffiltext{19}{Department of Astrophysical Science, Princeton University, NJ, USA}
\altaffiltext{20}{Department of Astronomy, The University of Tokyo, Tokyo, Japan}
\altaffiltext{21}{Institute of Space and Astronautical Science, Japan Aerospace Exploration Agency, Kanagawa, Japan}
\altaffiltext{22}{Hiroshima University, Hiroshima, Japan}
\altaffiltext{23}{Space Telescope Science Institute, MD, USA}
\altaffiltext{24}{National Meteorological Satellite Center, Chungbuk, Republic of Korea}
\altaffiltext{25}{Department of Astronomical Science, The Graduate University for Advanced Studies, Tokyo, Japan}
\altaffiltext{26}{Jet Propulsion Laboratory, California Institute of Technology, CA, USA}
\altaffiltext{27}{NASA Exoplanet Science Institute, California Institute of Technology, CA, USA}
\altaffiltext{28}{Institute of Astronomy and Astrophysics, Academia Sinica, Taipei, Taiwan}
\altaffiltext{29}{Swiss Federal Institute of Technology, Institute for Astronomy, Zurich, Switzerland}
\altaffiltext{30}{Kavli Institute for Physics and Mathematics of the Universe, The University of Tokyo, Chiba, Japan}
\altaffiltext{31}{Okayama University of Science, Okayama, Japan}
\altaffiltext{32}{H. L. Dodge Department of Physics and Astronomy, University of Oklahoma, OK, USA}


\maketitle

\begin{abstract}
We find a new substellar companion to the Pleiades member star, Pleiades~HII~3441, using the Subaru telescope with adaptive optics. The discovery is made as part of the high-contrast imaging survey to search for planetary-mass and substellar companions in the Pleiades and young moving groups. The companion has a projected separation of 0\farcs49~$\pm$~0\farcs02 (66~$\pm$~2~AU) and a mass of 68~$\pm$~5~$M_J$ based on three observations in the $J$-, $H$-, and $K_S$-band. The spectral type is estimated to be M7 ($\sim$2700~K), and thus no methane absorption is detected in the $H$ band. Our Pleiades observations result in the detection of two substellar companions including one previously reported among 20 observed Pleiades stars, and indicate that the fraction of substellar companions in the Pleiades is about 10.0$^{+26.1}_{-8.8}$\%. This is consistent with multiplicity studies of both the Pleiades stars and other open clusters.
\end{abstract}

\section{Introduction}\label{sec:intro}
The Pleiades has long been recognized as one of the nearest young open clusters (135~pc and 120~Myr, as discussed later). Young brown dwarfs have been extensively searched for in the Pleiades for studying the low-mass end of the initial mass function through various deep, wide-field imaging surveys (e.g., \cite{jameson+skillen_1989, stauffer+1989, stauffer+1998}). Among these studies, \citet{stauffer+1994} confirmed one and \citet{zapatero+1997} confirmed two brown dwarfs in the Pleiades. 

Adaptive optics imaging surveys are also a good tool for detection of faint companions such as brown dwarf and planetary-mass objects. \citet{bouvier+1997} directly imaged 144 G and K Pleiades members with the Canada-France-Hawaii telescope's adaptive optics system in order to investigate the stellar multiplicity, and found 22 binary systems and 3 triples with a separation between 0\farcs08--6\farcs9. They concluded that the binary fraction of the G- and K-type stars in the Pleiades (28\%~$\pm$~4\%) is similar to that of G-type field stars \citep{duquennoy+mayor_1991}. However, the stellar multiplicity is still uncertain with respect to the low-mass and closely bound objects, due to the limited sensitivity to faint companions with small separations from the primary and lack of follow-up, proper motion measurements. \citet{geisler+2012} and \citet{rodriguez+2012} each discovered a substellar mass companion around Pleiades~HII~1348 and HD~23514 with the adaptive optics system at the Keck observatory. 

We report in this paper the discovery of a new substellar mass companion to the Pleiades member star Pleiades~HII~3441 (2MASS~J03444394+2529574), with the Subaru high-contrast instrument HiCIAO (High Contrast Instrument for Subaru Next Generation Adaptive Optics; e.g., \cite{tamura+2006, suzuki+2010}) combined with the adaptive optics system AO188 (e.g., \cite{hayano+2008}) in the Subaru strategic program, the SEEDS (Strategic Exploration of Exoplanet and Disks with Subaru; \cite{tamura2009}) project. Table~\ref{tbl:primary} shows the properties of Pleiades~HII~3441. This object is a K-type star with membership probability between 78\% \citep{belikov+1998} and 98\% \citep{schilbach+1995} confirmed through both kinematic and photometric selection procedures. The portion of our Pleiades planet search was introduced in \citet{yamamoto+2013}. Pleiades~HII~3441B was confirmed after the conclusion of their survey. In addition, the companion status of unconfirmed companion candidates in \citet{yamamoto+2013} is substantiated in this paper, and new observations after that are introduced in the supplement.

We note that the employed distance and age of the Pleiades in this paper are 135~pc and 120~Myr, respectively, considering the previous works described below. The distance of the Pleiades had been stated to be between $\sim$120~pc (e.g., \cite{vanleeuwen_2009}) and 135~pc (e.g, \cite{soderblom+2005}). The controversy was settled by \citet{melis+2014}, and they showed the Pleiades to be at a distance of 136.2~$\pm$~1.2~pc. The age of the Pleiades ranges from 115 to 135~Myr based on various previous works (e.g., \cite{basri+1996, stauffer+1998a, barrado+2004, bell+2014}).

\begin{table}
    \tbl{Pleiades~HII~3441}{%
    \begin{tabular}{cl}
        \hline
        Other Names & 2MASS~J03444394+2529574\\
                    & TYC~1803-839-1\\
                    & SRS~71291, SSHJ~K121, PELS~41\\
        RA (h m s)  & 03 44 43.9\\
        DEC (d m s) & +25 29 57.1\\
        $J$         & 10.39~$\pm$~0.3~mag$^a$\\
        $H$         & 9.86~$\pm$~0.03~mag$^a$\\
        $K_S$       & 9.74~$\pm$~0.02~mag$^a$\\
        Proper Motion & (RA) 16.4~$\pm$~2.7~mas/yr$^b$\\
                    & (DEC) -48.8~$\pm$~2.6~mas/yr$^b$\\
        Spectral Type & K-type (K3~$\pm$~1)$^c$\\
        Membership  & Pleiades (78\%--98\%)\\
        \hline
    \end{tabular}} \label{tbl:primary}
\begin{tabnote}
 Note: (a) \cite{cutri+2003}. (b) \cite{hog+2000}. (c) The spectral sub-type was estimated using VOSA. See Appendix for the details.\\
\end{tabnote}
\end{table}

\section{Observations and Data Reduction}
We observed Pleiades~HII~3441 as part of the SEEDS survey. A summary of our target samples is shown in the supplementary material, including our target list (Supplementary Table~\ref{tab:targlist}), the observing logs (Supplementary Table~\ref{tab:obslog}), and the companion candidates list (Supplementary Table~\ref{tab:cclist}).

\subsection{Observations}
We observed Pleiades~HII~3441 using HiCIAO along with AO188 on the Subaru telescope. HiCIAO is a high contrast instrument for imaging exoplanets with a 2K$\times$2K HAWAII2-RG array and a pixel scale of 9.5~mas. Three observations were conducted for Pleiades~HII~3441, on 2011 September 4, on 2014 October 11, and on 2015 January 8, as shown in Table~\ref{tbl:obs}. In the 2011 observation, we used the simultaneous spectral differential imaging (SDI) mode in the $H$ band (e.g., see observations in \cite{janson+2013}), as well as the angular differential imaging (ADI) mode \citep{marois+2006}. The $H$ band is divided into two narrow bands ($H_S$: 1.486--1.628~$\mu$m and $H_L$: 1.643--1.788~$\mu$m) in the SDI mode, to search for methane absorption that is seen in low temperature objects ($<\sim$1300~K; \cite{sharp+burrows2007}). The ADI mode was used along with the SDI mode to achieve a high contrast at small angular separation. The other two observations were conducted as follow-up in order to measure the proper motion and photometric properties of the companion candidate. One was conducted in 2014, using the direct imaging (DI) + ADI mode in the $H$ band to check the proper motion of the companion candidate. Another was conducted in 2015, using the DI mode in the $J$, $H$, and $K_s$ bands to measure the colors. The rotation angle for the 2015 observations was small (see Table~\ref{tbl:obs}), because the ADI mode was used as replacement for dithering and not to achieve the highest possible contrast.

\begin{table*}
  \tbl{Observing Log of Pleiades~HII~3441}{%
  \begin{tabular}{ccccccc}
      \hline
      Obs. Date         & Mode      & Band          & Sub Exposure  & Coadd & Total Exposure    & Angle$^a$\\ 
      (UT)              &           &               & (s)           &       & (minutes)         & (degree)\\
      \hline
      2011 September 4  & SDI+ADI   & $H_S$, $H_L$  & 10        & 1     & 48.3          & 120.8\\
      2014 October 11   & DI+ADI    & $H$           & 1.5       & 30    & 15            & 75.9\\
      2015 January 8    & DI        & $H$           & 10        & 3, 10 & 4.7           & -\\
                        & DI        & $J$           & 20        & 1     & 4.7           & -\\
                        & DI+ADI$^b$ & $J$          & 20        & 1     & 13.7          & 18.6\\
                        & DI+ADI$^b$ & $K_s$        & 10        & 1     & 2.2           & 2.3\\
      \hline
    \end{tabular}}\label{tbl:obs}
\begin{tabnote}
Note: (a) Total rotational angle in case of the ADI mode. (b) ADI was used for dithering.
\end{tabnote}
\end{table*}

\subsection{Data Reduction}
In a pre-processing step, we conducted stripe removal, flat and dark corrections, bad pixels interpolation, and distortion correction according to the method described in \citet{yamamoto+2013}. For the ADI dataset, we used standard ADI reduction \citep{marois+2006} and Locally Optimized Combination of Imaging (LOCI; \cite{lafreniere+2007}). In the SDI mode, $H_S$ and $H_L$ images were simultaneously taken on the left and right half of the detector. We divided each frame into two, and then performed ADI reduction on each filter separately. For these analyses, we used Image Reduction and Analysis Facility (IRAF)- and Interactive Data Language (IDL)-based tools.

\section{Results and Discussion}
\begin{figure*}
 \begin{center}
  \includegraphics[width=\hsize,clip]{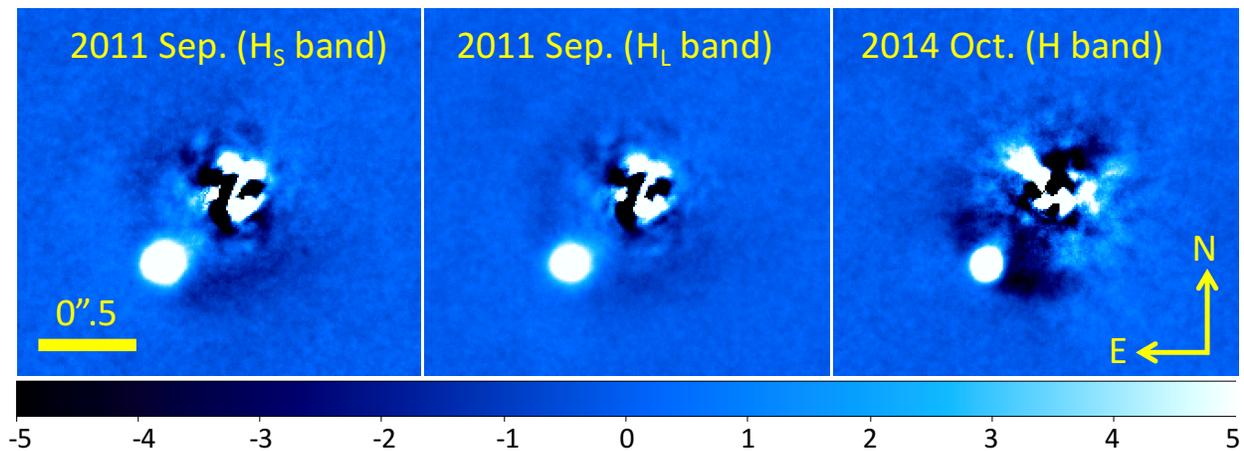} 
 \end{center}
\caption{Final Pleiades~HII~3441 images. (Left) reduced $H_S$-band image taken in the 2011 observation. (Middle) reduced $H_L$-band image taken in the 2011 observation. (Right) reduced $H$-band image taken in the 2014 observation. All images were analyzed using standard ADI. Pleiades~HII~3441B can be seen southeast of the primary star. There is no methane absorption in Pleiades~HII~3441B when left and middle panels are compared.}\label{fig:finimg}
\end{figure*}

Figure~\ref{fig:finimg} shows the final Pleiades~HII~3441 images taken in the 2011 and 2014 observations, reduced using standard ADI. Images reduced with LOCI were of similar sensitivity and are therefore not shown in addition. A companion candidate was detected southeast of the primary star, and subsequently confirmed as a companion object to the primary star.

\subsection{Proper Motion Check}
We measured the position of the companion candidate using the 2011, 2014, and 2015 observations (see Supplementary Table~\ref{tab:cclist}). Figure~\ref{fig:pm} shows the RA and DEC offsets of the companion candidate with the primary star at the origin. Pleiades~HII~3441 and its companion candidate are co-moving within 1$\sigma$ of the position errors, ruling out the possibility that the companion candidate is a background star by $>$~6$\sigma$. 

We also investigated the possibility that the companion candidate is another faint Pleiades member along the same line of sight, because the possibility cannot be ruled out completely due to the insufficient baseline of our observation to detect the orbital motion. The number of isolated Pleiades stars that could be chance aligned in the field of view of HiCIAO was estimated using the Pleiades stellar distribution (e.g., \cite{king+1962}) and luminosity function (e.g., \cite{bouvier+1998, jameson+2002, moraux+2003, bihain+2006}). The estimated number of stars with brightness similar to Pleiades~HII~3441B (15--16~mag in the $H$ band) is less than 0.03, when 21 Pleiades stars were observed. The likelihood of contamination is small, since the 21 targets are distant from the cluster center ($\sim$1\arcdeg).

We conclude that the companion candidate is indeed a bound companion to Pleiades~HII~3441 and refer to it in the following discussion as Pleiades~HII~3441B. The projected separation and position angle are shown in Table~\ref{tbl:properties} as 0\farcs49~$\pm$~0\farcs02 (66~$\pm$~2~AU) and 136.4$^\circ$~$\pm$~3.2$^\circ$, respectively. These values were derived by averaging all observations.

\begin{figure*}
 \begin{center}
  \includegraphics[width=\hsize,clip]{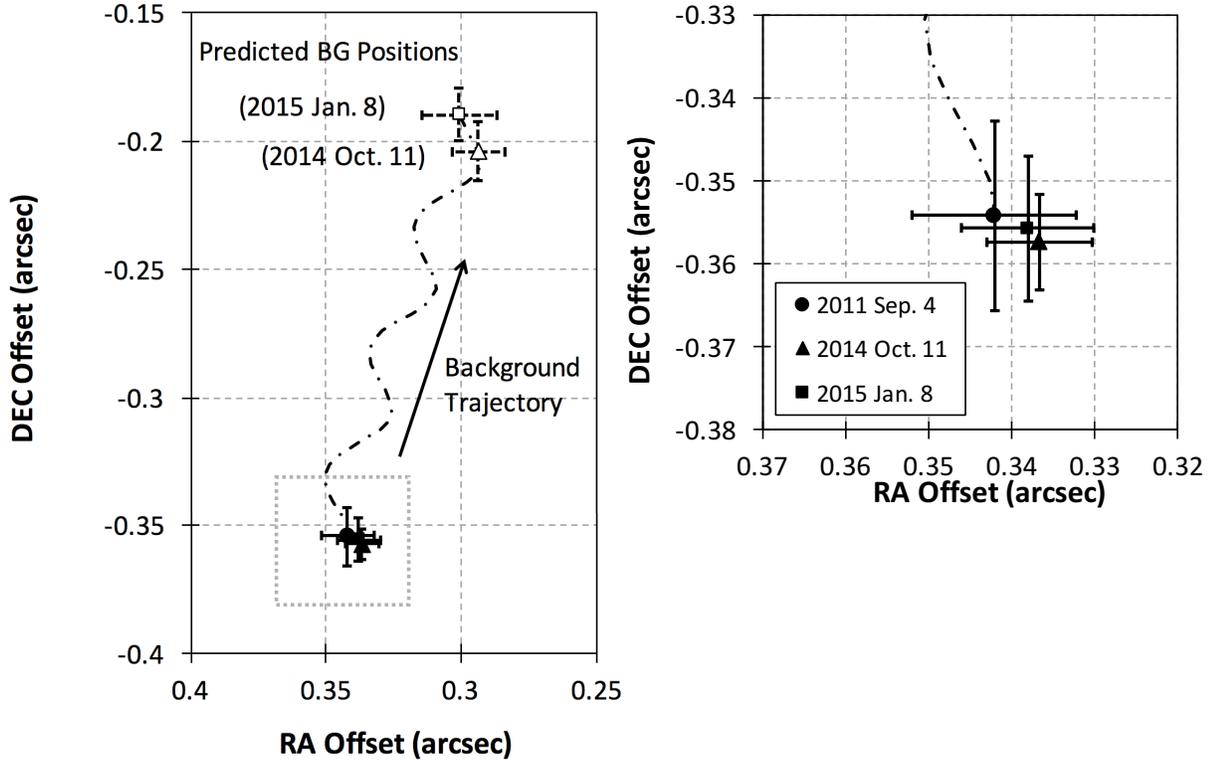} 
 \end{center}
\caption{(Left) Proper motion check of Pleiades~HII~3441B.  (Right) expanded view of the observed positions as marked by the gray box. Shown is the RA and DEC offset with respect to the primary star (Pleiades~HII~3441) at the origin. Filled marks indicate observed positions of the companion candidate. If the companion candidate were a background (BG) star, it would move along the dashed-dotted line. The predicted BG-case position in each epoch is shown as open marks. The BG trajectory only takes into account the proper motion of Pleiades~HII~3441 (see Table~\ref{tbl:primary}). Uncertainties are given as 1$\sigma$ interval.}\label{fig:pm}
\end{figure*}

\begin{table}
    \tbl{Properties of Pleiades~HII~3441B}{%
    \begin{tabular}{cl}
        \hline
        Projected Separation    & 0\farcs49~$\pm$~0\farcs02 (66~$\pm$~2~AU)$^a$ \\
        Position Angle          & 136.4$^\circ$~$\pm$~3.2$^\circ$\\
        $J$                     & 15.64~$\pm$~0.14~mag ($68~\pm~3~M_J$)$^b$\\
        $H$                     & 15.23~$\pm$~0.08~mag ($65~\pm~2~M_J$)$^b$\\
        $K_S$                   & 14.71~$\pm$~0.06~mag ($72~\pm~2~M_J$)$^b$\\
        $\Delta H_S$            & 5.47~$\pm$~0.19~mag\\
        $\Delta H_L$            & 5.29~$\pm$~0.29~mag\\
        $H_S-H_L$               & 0.18~$\pm$~0.26~mag\\
        \hline
    \end{tabular}} \label{tbl:properties}
\begin{tabnote}
 Note: $\Delta H_S$ and $\Delta H_L$ are relative magnitudes to the primary star. (a) Projected separation is calculated using the Pleiades distance 135~pc. (b) Mass at 120~Myr is estimated using the BT-Settl model. 
\end{tabnote}
\end{table}

\subsection{Multiband Photometry} \label{sec:multiphoto}
\begin{figure*}
 \begin{center}
    \includegraphics[width=\hsize,clip]{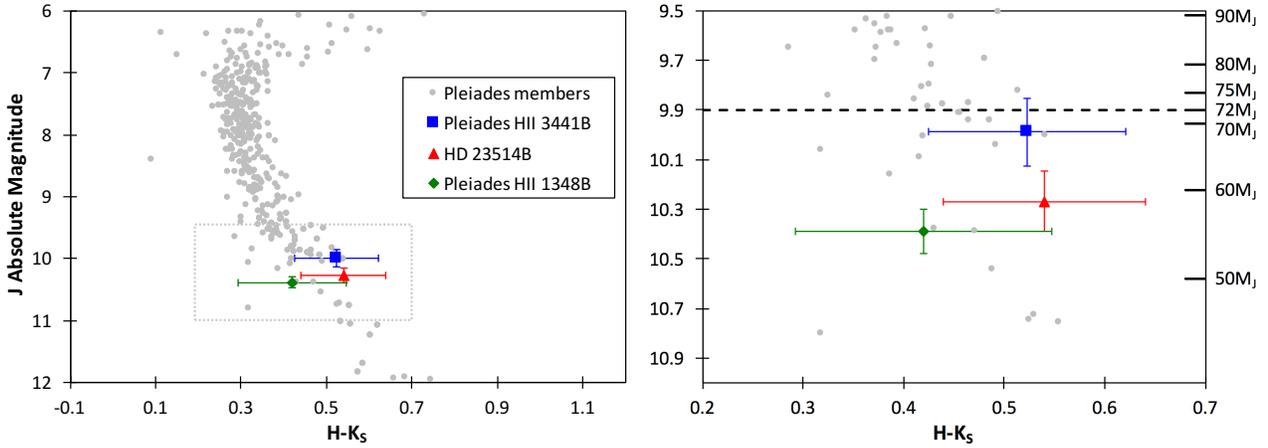}
 \end{center}
 \caption{Color-magnitude diagram of Pleiades substellar companions. Shown is absolute $J$-band magnitude over $H-K_S$ color. (Left) Pleiades member stars and three imaged substellar companions. (Right) expanded view of the three companions' color as marked by the grey box. Filled square, triangle, and diamond marks indicate Pleiades~HII~3441B, HD~23514B \citep{rodriguez+2012}, and Pleiades~HII~1348B \citep{geisler+2012}, respectively. Small gray circles are low-mass Pleiades members taken from \citet{lodieu+2007}. The mass scale is shown on the right axis of the right panel. The BT-Settl model is used for a magnitude-to-mass conversion. The hydrogen-burning limit (72~$M_J$) is also shown as dashed line in the right panel.} \label{fig:color}
\end{figure*}

The ADI-reduced images typically suffer from the self-subtraction which biases the brightness measurements if not taken into account. To avoid effects caused by the self-subtraction, we used each image before the ADI reduction for photometry. We performed aperture photometry using the $apphot$ task in IRAF, with an 8~pixels-radius aperture corresponding to the FWHM of the point spread function. The halo of the primary star affects the accuracy of the photometry of Pleiades~HII~3441B due to its small angular separation. Therefore, we estimated the offset value contributed by the halo at the position of Pleiades~HII~3441B, assuming the halo is azimuthally symmetric. The offset value is calculated in an annulus at the same angular separation as HII~3441B and centered on the primary star. We performed relative photometry to the primary star based on its infrared magnitudes reported in \citet{cutri+2003}, in order to obtain the brightness of Pleiades~HII~3441B. For data taken in 2015, the $J$-, $H$-, and $K_S$-band magnitudes are 15.64~$\pm$~0.14, 15.23~$\pm$~0.08, and 14.71~$\pm$~0.06, respectively (see Table~\ref{tbl:properties}). The $J$-band magnitude corresponds to a mass of $68~\pm~3~M_J$ and a temperature of $\sim$2700~K at 120~Myr according to the BT-Settl model \citep{baraffe+2015}. The mass is below the hydrogen-burning limit (72~$M_J$: e.g., \cite{chabrier+2000}), and therefore Pleiades~HII~3441B classifies as a brown dwarf. The spectral type is estimated to be M7 from the photometry-derived temperature \citep{pecaut+mamajek+2013}. Figure~\ref{fig:color} shows a color-magnitude diagram of three substellar companions in the Pleiades that have been reported (HD~23514B; \cite{rodriguez+2012} and Pleiades~HII~1348B; \cite{geisler+2012}). The properties of Pleiades~HII~3441B are consistent with the other two substellar companions and substellar members in the Pleiades reported in \citet{lodieu+2007}. 

While the mass estimate derived from the $J$- and $H$-band magnitudes (68~$\pm$~3~$M_J$, 65~$\pm$~2~$M_J$) falls into the highest mass brown dwarf regime, the mass derived from the $K_s$-band magnitude (72~$\pm$~2~$M_J$) potentially crosses the hydrogen-burning limit. We note that this is an object that is close to the boundary between the stellar and substellar regime.

\subsection{Methane Absorption}
We tested for the existence of methane absorption using the SDI observation. The aperture photometry was performed using the same procedures as described in Section~\ref{sec:multiphoto}, and the results are shown in Table~\ref{tbl:properties}. We obtained an $H_S-H_L$ color of 0.18~$\pm$~0.26, which means that there is no methane absorption in the atmosphere of Pleiades~HII~3441B. This is consistent with our conclusion about the object being an M7-type dwarf ($\sim$2700~K), because methane is considered to condense below $\sim$1300~K \citep{sharp+burrows2007}. We also compared its color with theoretical values derived from simulated BT-Settl spectra generated using the PHOENIX web simulator\footnote{https://phoenix.ens-lyon.fr/simulator/index.faces} and assuming stellar parameters of Pleiades~HII~3441B (the effective temperature of 2700~K, logarithm of the surface gravity of 5.0, and solar metallicity). The theoretical $H_S-H_L$ color is 0.04 and is consistent with our measurements within 1$\sigma$. There is no contradiction between Pleiades~HII~3441B and a late M-dwarf spectrum.

\subsection{Discussion: Fraction of Substellar Companions in the Pleiades}
As introduced in Section~\ref{sec:intro}, the Pleiades multiplicity is well investigated except for low-mass and close companions.  Deriving a fraction of close substellar companions is therefore an essential key for the determination of the initial mass function, and might help to understand the formation mechanisms in the cluster. 

We detected three substellar companions and a stellar companion among 21 observed stars in the Pleiades including two companions reported in \citet{yamamoto+2013} (see also Supplementary Table~\ref{tab:cclist}). To discuss an unbiased sample, we consider 20 stars after removing Pleiades~HII~1348 since we intended to observe its companion for the comparison with the previous work \citep{geisler+2012} for this source. The binary fraction of $3/20~=~15\%$ has a strong bias, because known binaries have been removed from the target sample (see \cite{yamamoto+2013}). Therefore, we focus on substellar companions to the Pleiades stars. The fraction of detected substellar companions is $2/20~=~10\%$ in our survey. If we assume Poisson statistics and the 95\% confidence interval, the fraction ranges from 1.2\% to 36.1\%. The companion mass and projected separation range that our survey probed spans roughly from 10 to 75~$M_J$ and from 65 to $\sim$1000~AU, respectively, according to detection limits reported in \citet{yamamoto+2013}. 

\begin{table}
    \tbl{Fractions of Substellar Companions}{%
    \begin{tabular}{ccl}
        \hline
        \citet{bouvier+1997}            &   Pleiades        & 4.9--6.1\%\\
        \citet{richichi+2012}           &   Pleiades        & 5.9\% (re-calculated)\\
        \hline
        \citet{patience+2002}           & $\alpha$~Persei   & 9\%\\
        \citet{patience+2002}           &   Praesepe        & 10\%\\
        \hline
        \citet{metchev+hillenbrand2009} & including field stars & 3.2\%\\
        \citet{brandt+2014}             & including field stars & 1--3.1\%\\
        \hline
        \hline
        This Work                       &   Pleiades        & 10.0$^{+26.1}_{-8.8}$\% (95\% confidence interval)\\
        \hline
    \end{tabular}} \label{tbl:fraction}
\begin{tabnote}
 
\end{tabnote}
\end{table}

We compare the fraction with those of the other works shown in Table~\ref{tbl:fraction}. \citet{bouvier+1997} found that the fraction at a mass ratio of $<$~0.1 is 4.9\% (1\farcs0--2\farcs0) and 6.1\% (2\farcs0--6\farcs9) using 144 Pleiades stars. Recent studies also reported similar fractions. For example, using only substellar companions (absolute $J>$~10 mag) among 17 observed stars reported in \citet{richichi+2012}, a fraction of 5.9\% is obtained. Small sample high-contrast-imaging observations conducted in the Pleiades resulted in no detection of substellar companions within their 10 samples \citep{itoh+2011}. In other open clusters, we only note that a large imaging survey revealed that the multiplicity with a mass ratio of $>$~0.25 and a projected separation range of 26 to 581~AU is 9\% and 10\% using 142 $\alpha$~Persei stars and 100 Praesepe stars, respectively \citep{patience+2002}. Our fraction is consistent with those of the previous studies, although the parameter space explored by those previous surveys is not the same as in our observations. In addition, we point out that there is no large deviation of the fraction among open clusters, which has already been reported in \citet{duchene+kraus2013}.

Recently, several works have derived the fractions of substellar companions which included young field stars. \citet{metchev+hillenbrand2009} revealed that the fraction is 3.2\% using 266 solar-type stars. The SEEDS results also support this with a fraction of 1--3.1\% using 250 high-contrast-imaging stars \citep{brandt+2014}. These studies included field stars and derived smaller fractions than ours using Bayesian statistics. This might be due to the difference in sample size (insufficient sample size for open cluster stars) and the estimation method, because these fractions are roughly within our large uncertainty. A much larger survey of the Pleiades would be needed to draw general conclusion on the multiplicity differences between open clusters and field star populations.

\section{Summary}
We discovered a substellar companion to Pleiades~HII~3441 as part of the SEEDS survey using Subaru/HiCIAO together with AO188. Pleiades~HII~3441B has a separation of 0\farcs49~$\pm$~0\farcs02 (66~$\pm$~2~AU) and a mass of 68~$\pm$~5~$M_J$, based on $J$-, $H$-, $K_S$-band observations, that is below but very close to the hydrogen-burning limit. We also confirmed two previously known companions among the 21 observed Pleiades stars reported by \citet{yamamoto+2013}. After an object which was observed for comparison with the previous study was removed from our sample, the fraction of substellar companion detections is 10$^{+26.1}_{-8.8}$\% if we assume Poisson statistics and the 95\% confidence interval. Our result is consistent with previous studies of the Pleiades and other clusters.

\begin{ack}
The authors recognize and acknowledge the significant cultural role and reverence that the summit of Mauna Kea has always had within the indigenous Hawaiian community. We are most fortunate to have the opportunity to conduct observations from this mountain. We thank the anonymous referee for careful reading our manuscript and for giving helpful comments. The authors are grateful for David Lafreni\'{e}re for generously providing the source code for the LOCI algorithm. This publication makes use of VOSA to estimate the primary spectral type, developed under the Spanish Virtual Observatory project supported from the Spanish MICINN through grant AyA2011-24052. This work was partially supported by the Grant-in-Aid for JSPS fellows (Grant Number 25-8826). J.C. was supported by the U.S. National Science Foundation under Award No. 1009203.
\end{ack}

\appendix
\section*{Estimation of Pleiades~HII~3441 Spectral Type} \label{ap:sed}
\begin{figure}
 \begin{center}
    \includegraphics[width=\hsize,clip]{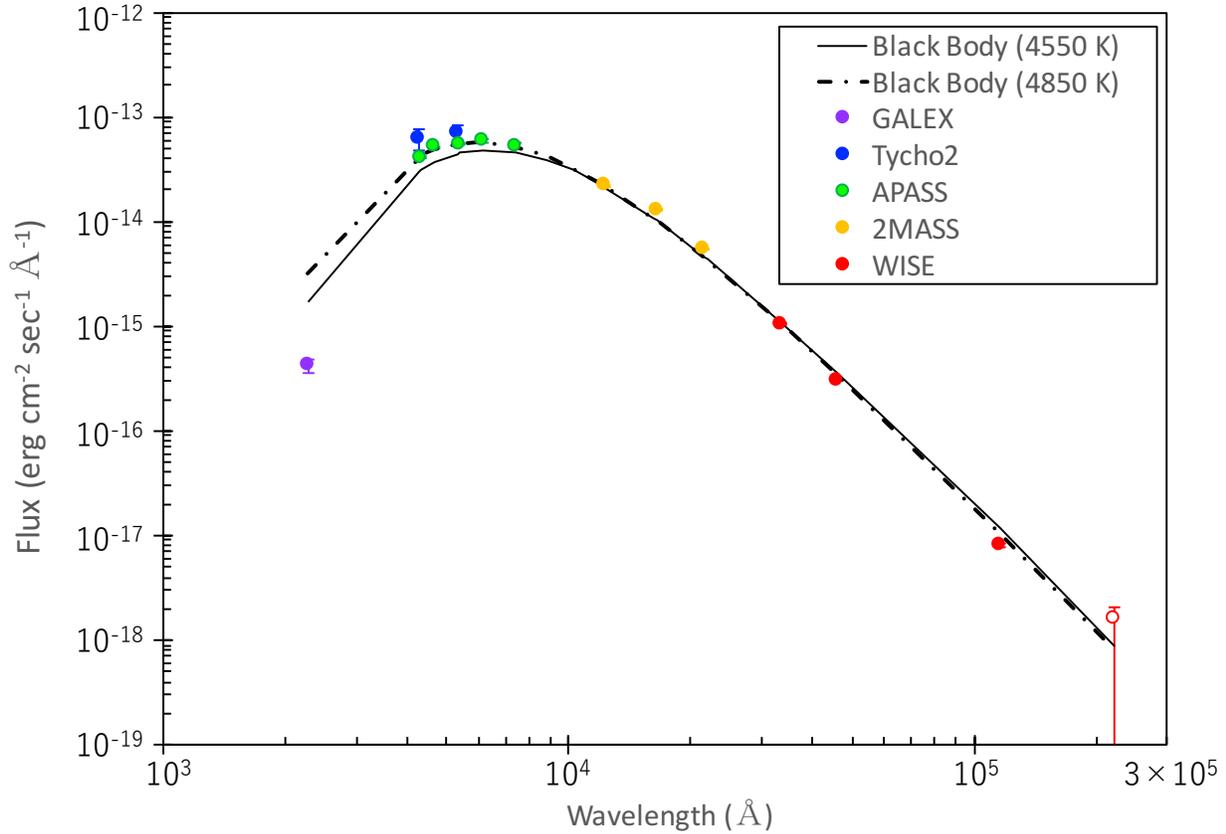}
 \end{center}
 \caption{SED of Pliades~HII~3441 generated by VOSA. Black lines indicate the 4550~K and 4850~K black bodies that are most fitted model using the data with and without GALEX, respectively. Circles are published photometric results in the literature \citep{bianchi+2000, hog+2000, cutri+2003, henden+2009, wright+2010}. Open mark was not used for SED fitting because it is upper limit.} \label{fig:sed}
\end{figure}

The spectral type of the primary star has been known as K-type, but has not been investigated in more detail. We therefore constructed the spectral energy distribution (SED) using published photometric results in the literature, and estimated the spectral type by fitting the black body spectrum. The flux at each wavelength is taken from GALEX (Galaxy Evolution Explorer; \cite{bianchi+2000}), APASS (the AAVSO Photometric All-Sky Survey; \cite{henden+2009}), Tycho2 (Tycho-2 Catalogue; \cite{hog+2000}), 2MASS (Two Micron All-Sky Survey; \cite{cutri+2003}), and WISE (Wide-field Infrared Survey Explorer; \cite{wright+2010}) catalogues. The black body spectrum was fitted by $\chi^2$ minimization using the VOSA web tool\footnote{http://svo2.cab.inta-csic.es/theory/vosa/}. The fitted SED is shown in Figure~\ref{fig:sed}. The effective temperature was estimated as 4550~$\pm$~10~K using all data, and as 4850~$\pm$~20~K without GALEX data. 
For an additional level of characterization, we estimated the spectral type and effective temperature of the star by interpolating its optical and near-IR photometric colors within the main-sequence color-temperature conversion table of \citet{pecaut+mamajek+2013}. We find an effective temperature of 4800~$\pm$~160~K. The uncertainties were calculated using Monte Carlo methods. This temperature is consistent with the SED determination. We concluded that Pleiades~HII~3441 has an effective temperature of 4700~$\pm$~200~K, and corresponds to a spectral type of K3~$\pm$~1 \citep{pecaut+mamajek+2013}.

\newpage
\input{supplement.tex}
\end{document}

%% file: supplement.tex
\renewcommand{\tablename}{Supplementary~Table}
\setcounter{table}{0}

\section*{Supplement: Our Survey (Subsamples of SEEDS observations)}\label{sp:seeds}
Our survey is conducted as part of the SEEDS program. The SEEDS survey results have been reported in numerous works (e.g., \cite{janson+2013b, brandt+2014b, uyama+2016}). The supplement contains results of subsamples that we obtained, 
including previously reported results of Pleiades stars \citep{yamamoto+2013}. Supplementary~Table~\ref{tab:targlist} represents our target list. We focused on young stars whose ages are well known. Each target belongs to one of the moving groups (MG) or associations: the Pleiades (120~Myr, 135~pc), Ursa Major MG (500~Myr, $\sim$25~pc; \cite{king+2003}), Octans-Near association ($<$100~Myr, $\sim$90~pc; \cite{zuckerman+2013}), and AB Doradus MG ($\sim$50~Myr, 30~pc; \cite{lopez+2006}). One target (GJ~212) is considered to belong to either Hercules-Lyra association ($\sim$200~Myr, $<$25~pc; \cite{lopez+2006}) or Local association (20--150~Myr: \cite{montes+2001}). Supplementary~Table~\ref{tab:obslog} shows the observing logs of data taken after September 2012, excluding the ones reported in \citep{yamamoto+2013}. Supplementary~Table~\ref{tab:cclist} is the companion candidates list including their projected separation, position angle (PA), and photometric results. The companion status is determined by confirmation of their proper motion.

\input{s_table1.tex}

\input{s_table2.tex}

\input{s_table3.tex}

%% file: s_table1.tex
\begin{longtable}{ccccccccccc}
 \caption{Target List} \label{tab:targlist}
    \hline
    Name & HD Number & RA & DEC & Sp.T & Group & Distance & $R$ & $J$ & $H$ & $K_S$\\
     & & (h m s) & (d m s) & & & (pc) & (mag) & (mag) & (mag) & (mag)\\
    \hline
 \endfirsthead
    \hline
    Name & HD Number & RA & DEC & Sp.T & Group & Distance & $R$ & $J$ & $H$ & $K_S$\\
     & & (h m s) & (d m s) & & & (pc) & (mag) & (mag) & (mag) & (mag)\\
    \hline
 \endhead
    \hline
 \endfoot
    \hline
    \multicolumn{11}{l}{Note: $\dagger$ Previously reported in \citet{yamamoto+2013}.}\\
    \multicolumn{11}{l}{Group column abbreviations}\\
    \multicolumn{11}{l}{  AB Dor: AB Doradus moving group \citep{lopez+2006}}\\
    \multicolumn{11}{l}{  UMa: Ursa Major moving group \citep{king+2003}}\\
    \multicolumn{11}{l}{  Pleiades: Pleiades open cluster \citep{yamamoto+2013}}\\
    \multicolumn{11}{l}{  HLA: Hercules-Lyra moving group \citep{lopez+2006}}\\
    \multicolumn{11}{l}{  LA: Local Association \citep{montes+2001}}\\
    \multicolumn{11}{l}{  ONA: Octans-Near Association \citep{zuckerman+2013}}\\
    \multicolumn{11}{l}{$R$ magnitude: NOMAD database \citep{zacharias+2005}}\\
    \multicolumn{11}{l}{$J$, $H$, $K_S$ magnitudes: 2MASS \citep{cutri+2003}}\\
    \multicolumn{11}{l}{Other properties are taken from the above group citations.}\\
 \endlastfoot
    HIP~6276            & -     & 01 20 32.3 & -11 28 03.7 & G9 & AB Dor    & 35.1 & 7.90 & 7.03 & 6.65 & 6.55\\
    Chi~Cet~B           & 11131 & 01 49 23.4 & -10 42 12.9 & G1 & UMa       & 23.0 & 6.39 & 5.54 & 5.29 & 5.15\\
    HD~23061$^\dagger$  & 23061 & 03 42 55.1 & +24 29 35.1 & F5 & Pleiades  & 135  & 9.01 & 8.51 & 8.33 & 8.26\\
    HII~3456$^\dagger$  & -     & 03 43 27.1 & +25 23 15.3 & G2 & Pleiades  & 135  & 11.21 & 9.75 & 9.31 & 9.22\\
    HD~23247$^\dagger$  & 23247 & 03 44 23.5 & +24 07 57.6 & F3 & Pleiades  & 135  & 8.85 & 8.08 & 7.81 & 7.77 \\
    HII~3441            & -     & 03 44 43.9 & +25 29 57.1 & K  & Pleiades  & 135  & 11.41 & 10.39 & 9.86 & 9.74\\
    HII~636$^\dagger$   & -     & 03 45 22.2 & +23 28 18.2 & K  & Pleiades  & 135  & 11.69 & 10.47 & 9.96 & 9.85\\
    V855~Tau$^\dagger$  & -     & 03 45 40.2 & +24 37 38.1 & F8 & Pleiades  & 135  & 9.37  & 8.62 & 8.34 & 8.29\\
    BD+23~514$^\dagger$ & -     & 03 45 41.9 & +24 25 53.5 & G5 & Pleiades  & 135  & 11.15 & 9.80 & 9.53 & 9.40\\
    V1171~Tau$^\dagger$ & -     & 03 46 28.4 & +24 26 02.1 & G8 & Pleiades  & 135  & 10.51 & 9.64 & 9.27 & 9.16\\
    HD~23514$^\dagger$  & 23514 & 03 46 38.4 & +22 55 11.2 & G0 & Pleiades  & 135  & 8.96 & 8.48 & 8.29 & 8.15\\
    HD~282954$^\dagger$ & 282954& 03 46 38.8 & +24 57 34.7 & G0 & Pleiades  & 135  & 9.98 & 9.11 & 8.85 & 8.76\\
    HII~1348$^\dagger$  & -     & 03 47 18.1 & +24 23 26.8 & K5 & Pleiades  & 135  & 12.43 & 10.49 & 9.83 & 9.72\\
    TYC~1800-2144-1$^\dagger$ & - & 03 48 34.5 & +23 26 05.3 & G0 & Pleiades & 135 & 10.37 & 9.20 & 8.98 & 8.87\\
    HD~23863$^\dagger$  & 23863 & 03 49 12.2 & +23 53 12.5 & A7 & Pleiades  & 135  & 7.98 & 7.67 & 7.60 & 7.58\\
    HII~2311$^\dagger$  & -     & 03 49 28.7 & +23 42 44.1 & G2 & Pleiades  & 135  & 10.91 & 9.91 & 9.54 & 9.43\\
    HD~23912$^\dagger$  & 23912 & 03 49 32.7 & +23 22 49.5 & F3 & Pleiades  & 135  & 8.88 & 8.26 & 8.10 & 8.04\\
    HII~2366$^\dagger$  & -     & 03 49 36.5 & +24 17 46.1 & G2 & Pleiades  & 135  & 10.88 & 10.03 & 9.63 & 9.55\\
    HII~2462$^\dagger$  & -     & 03 49 50.4 & +23 42 20.2 & G2 & Pleiades  & 135  & 11.50 & 10.07 & 9.70 & 9.60\\
    BD+22~574$^\dagger$ & -     & 03 49 56.5 & +23 13 07.0 & F8 & Pleiades  & 135  & 10.02 & 9.15 & 8.85 & 8.80\\
    V1174~Tau$^\dagger$ & -     & 03 50 34.6 & +24 30 28.2 & G8 & Pleiades  & 135  & 11.61 & 10.70 & 10.20 & 10.08\\
    HD~24132$^\dagger$  & 24132 & 03 51 27.2 & +24 31 07.1 & F2 & Pleiades  & 135  & 8.49 & 8.06 & 7.93 & 7.88\\
    V1054~Tau$^\dagger$ & -     & 03 51 39.3 & +24 32 56.1 & K  & Pleiades  & 135  & 12.35 & 10.42 & 9.92 & 9.81\\
    BD-01~565           & 24916 & 03 57 28.7 & -01 09 34.1 & K4 & UMa       & 15.8 & 7.34 & 6.06 & 5.49 & 5.34\\
    ome~Tau             & 27045 & 04 17 15.7 & +20 34 42.9 & A3 & ONA       & 29 & 4.77 & 4.79 & 4.58 & 4.36\\
    GJ~212              & 233153& 05 41 30.7 & +53 29 23.3 & M0 & HLA/LA    & 12.5 & 8.81 & 6.59 & 5.96 & 5.76\\
    V1386~Ori           & 41593 & 06 06 40.5 & +15 32 31.6 & K0 & UMa       & 15.5 & 6.24 & 5.32 & 4.94 & 4.82\\
    HIP~36624           & 59507 & 07 31 55.6 & +38 53 45.8 & A2 & ONA       & 80.4 & 6.53 & 6.41 & 6.47 & 6.42\\
    CCDM~J08316+3458A   & 71974 & 08 31 35.0 & +34 57 58.4 & G5 & UMa       & 28.7 & 6.82 & 5.92 & 5.50 & 5.47\\
    BD-13~2855          & 81659 & 09 26 42.8 & -14 29 26.7 & G6 & UMa       & 39.9 & 7.44 & 6.69 & 6.41 & 6.31\\
    DS~Leo              & 95650 & 11 02 38.3 & +21 58 01.7 & M2 & UMa       & 11.7 & 8.64 & 6.52 & 5.90 & 5.69\\
    BD+52~1638          & 109647& 12 35 51.3 & +51 13 17.3 & K0 & UMa       & 26.3 & 7.94 & 6.73 & 6.25 & 6.16\\
    HR~4803             & 109799& 12 37 42.3 & -27 08 20.0 & F1 & UMa       & 34.6 & 5.23 & 4.76 & 4.63 & 4.54\\
    BD+22~2522          & 112196& 12 54 40.0 & +22 06 28.6 & F8 & UMa       & 34.3 & 6.67 & 5.88 & 5.63 & 5.55\\
    GJ~516              & -     & 13 32 44.8 & +16 48 40.9 & M3 & UMa       & 13.8 & 12.04 & 7.64 & 7.07 & 6.83\\
    HR~5148             & 119124& 13 40 23.2 & +50 31 09.9 & F8 & ONA       & 25 & 6.02 & 5.28 & 5.11 & 5.02\\
    GJ~9457B            &119124B& 13 40 24.5 & +50 30 57.6 & K7 & ONA       & 25 & 9.98 & 7.79 & 7.16 & 7.01\\
    HR~7451             & 184960& 19 34 19.8 & +51 14 11.8 & F7 & UMa       & 25.6 & 5.43 & 4.70 & 4.59 & 4.49\\
    BD-00~4333          & 211575& 22 18 04.3 & -00 14 15.6 & F3 & UMa       & 41.5 & 6.12 & 5.59 & 5.35 & 5.33\\
\end{longtable}

%% file: s_table2.tex
\begin{longtable}{ccccccc}
  \caption{Observing Logs Taken after September 2012} \label{tab:obslog}
    \hline
    Name & Date & Mode & Filter & Sub Exposure & Coadd & Total Exposure\\
     & (UT) & & & (s) & & (minutes)\\
    \hline
  \endfirsthead
    \hline
    Name & Date & Mode & Filter & Sub Exposure & Coadd & Total Exposure\\
     & (UT) & & & (s) & & (minutes)\\
    \hline
  \endhead
   \hline
  \endfoot
   \hline
   \multicolumn{7}{l}{Note: $^{\ast}$ Extremely poor condition}\\
  \endlastfoot
    HIP~6276    & 2012 Nov. 7       & DI+ADI    & $H$           & 15    & 10    & 40\\
    Chi~Cet~B   & 2013 Jan. 3       & DI+ADI    & $H$           & 5     & 10    & 31.7\\
    HII~3441    & 2011 Sep. 4       & SDI+ADI   & $H_S$, $H_L$  & 10    & 1     & 48.3\\
                & 2014 Oct. 11      & DI+ADI    & $H$           & 1.5   & 30    & 15\\
                & 2015 Jan. 8       & DI        & $J$, $H$, $K_S$ & 20, 10, 10 & 1, 3/10, 1 & 18.4, 4.7, 2.2\\
    V1174~Tau   & 2012 Sep. 12      & DI        & $H$           & 10    & 3     & 25\\
                & 2013 Oct. 16      & DI        & $J$, $H$, $K_S$ & 30, 30, 30 & 1, 1, 1 & 8, 9, 8\\
                & 2013 Nov. 24      & DI+ADI    & $H$           & 10, 1.5 & 5, 50 & 28.8\\
                & 2014 Oct. 7, 9    & DI        & $H$           & 60, 30 & 1, 1 & 15, 25\\
    V1054~Tau   & 2012 Sep. 11, 12  & DI        & $H$           & 10, 10 & 1, 3 & 8.5, 10\\
                & 2013 Feb. 26$^{\ast}$ & DI    & $H$           & 20    & 3     & 36\\
    BD-01~565   & 2012 Nov. 5       & DI+ADI    & $H$           & 1.5   & 10    & 13\\
    ome~Tau     & 2014 Jan. 20      & DI+ADI    & $H$           & 1.5   & 10    & 17.5\\
                & 2014 Oct. 10      & DI        & $H$           & 30    & 1     & 12.5\\
    GJ~212      & 2013 Jan. 2       & DI+ADI    & $H$           & 15    & 3     & 43.5\\
                & 2013 Oct. 17      & DI        & $H$           & 15    & 10    & 15\\
    V1386~Ori   & 2013 Jan. 3$^{\ast}$ & DI+ADI & $H$           & 10    & 3     & 61\\
    HIP~36624   & 2013 Nov. 23      & DI+ADI    & $H$           & 10    & 10    & 48.3\\
                & 2015 Jan. 7       &  DI, DI+ADI & $H$         & 60, 20 & 1, 1 & 50, 21.3\\
    CCDM~J08316+3458A & 2013 Nov. 24 & DI+ADI   & $H$           & 10    & 10    & 45\\
                & 2015 Jan. 8       & DI+ADI    & $H$           & 10    & 10    & 38.3\\
    BD-13~2855  & 2013 Jan. 1       & DI+ADI    & $H$           & 15    & 3     & 33.8\\
                & 2014 Jan. 20      & DI+ADI    & $H$           & 15    & 2     & 13\\
                & 2014 Apr. 23      & DI        & $J$, $H$      & 60, 30 & 1, 1 & 13, 10\\
    DS~Leo      & 2014 Jan. 21      & DI+ADI    & $H$           & 1.5   & 10    & 17.3\\
    BD+52~1638  & 2013 Feb. 26      & DI+ADI    & $H$           & 20    & 3     & 33\\
    HR~4803     & 2013 Jan. 2       & DI+ADI    & $H$           & 10    & 10    & 51.7\\
    BD+22~2522  & 2013 May 18       & DI+ADI    & $H$           & 5     & 10    & 33.3\\
    GJ~516      & 2014 Jun. 8       & DI+ADI    & $H$           & 5     & 10    & 21.7\\
    HR~5148     & 2014 Apr. 23      & DI+ADI    & $H$           & 1.5   & 10    & 26.3\\
    GJ~9457B    & 2014 Jun. 7       & DI+ADI    & $H$           & 15    & 3     & 33.7\\
    HR~7451     & 2012 Sep. 14      & DI+ADI    & $H$           & 1.5   & 10    & 19\\
                & 2014 Apr. 24      & DI        & $H$           & 60    & 1     & 31\\
    BD-00~4333  & 2012 Nov. 5       & DI+ADI    & $H$           & 1.5   & 10    & 20.8\\
                & 2013 Oct. 16$^{\ast}$ & DI        & $K_S$         & 10    & 1     & 5.3\\
                & 2014 Jun. 7       & DI        & $J$, $H$, $K_S$ & 60, 30, 60 & 1, 1, 1 & 3, 10, 2\\
\end{longtable}

%% file: s_table3.tex
\begin{longtable}{ccccccc}
 \caption{Companion Candidates List} \label{tab:cclist}
    \hline
    Primary Name & CC Number & Status & Date & Separation & PA & $H$\\
     & & & (UT) & (arcsec) & (degree) & (mag) \\
    \hline
 \endfirsthead
    \hline
    Primary Name & CC Number & Status & Date & Separation & PA & $H$\\
     & & & (UT) & (arcsec) & (degree) & (mag) \\
    \hline
 \endhead
    \hline
 \endfoot
    \hline
    \multicolumn{7}{l}{Note: Status column abbreviations}\\
    \multicolumn{7}{l}{ BG: background objects, FG: foreground objects, C: companions, U: undefined objects}\\
    \multicolumn{7}{l}{Typical uncertainties of separation, PA, and photometric results are 0\farcs01--0\farcs03, 0.1--1.0$^{\circ}$, and 0.1--0.9~mag.}\\
    \multicolumn{7}{l}{$\dagger$ Previously reported in \citet{yamamoto+2013}, and only the latest properties are shown.}\\
    \multicolumn{7}{l}{$\ddagger$ \citet{yamamoto+2013} could not conclude the companion status of these objects.}\\
    \multicolumn{7}{l}{$\ast$ Data were not suitable for accurate measurements due to poor condition.}\\
    \multicolumn{7}{l}{$\diamond$ This was judged by comparison with the previous work \citep{ammler+2016}.}\\
    \multicolumn{7}{l}{$\star$ This CC was not detected in the second epoch.}\\
    \multicolumn{7}{l}{$\sharp$ This object locates very close to CC2. It has a possibility of artifacts.}\\
 \endlastfoot
    Chi~Cet~B   & CC1               & C (known)     & 2013 Jan. 3       & 0.278 & 349.7 & 8.68\\
                & CC2               & maybe C$^\diamond$ & 2013 Jan. 3  & 6.15  & 353.4 & 9.39\\
    HD~23247    & CC1$^{\ddagger}$  & C (stellar)   & 2011 Jan. 27      & 3.86  & 267.2 & 11.0\\
                &                   &               & 2011 Dec. 24      & 3.83  & 267.0 & -\\
    HII~3441    & CC1               & C             & 2011 Sep. 4       & 0.493 & 135.9 & -\\
                &                   &               & 2014 Oct. 11      & 0.491 & 136.9 & 15.2\\
                &                   &               & 2015 Jan. 8       & 0.491 & 136.5 & -\\
    V855~Tau    & CC1$^\dagger$     & maybe FG      & 2011 Jan. 28      & 8.05  & 19.5  & 17.2\\
    V1171~Tau   & CC1$^\dagger$     & BG            & 2012 Dec. 31      & 12.52 & 134.6 & 17.8\\
                & CC2$^\dagger$     & BG            & 2012 Dec. 31      & 12.63 & 135.5 & 18.5\\
                & CC3$^{\ddagger}$  & BG            & 2009 Nov. 1       & 9.08  & 125.9 & 19.0\\
                &                   &               & 2012 Dec. 31      & 8.94  & 125.4 & -\\
    HD~23514    & CC1$^\dagger$     & C             & 2010 Dec. 1       & 2.65  & 227.6 & 15.4\\
    HD~282954   & CC1$^\dagger$     & BG            & 2012 Sep. 12      & 8.94  & 103.3 & 14.4\\
    HII~1348    & CC1$^\dagger$     & C             & 2011 Dec. 23      & 1.12  & 346.1 & 15.7\\
    HD~23912    & CC1$^\dagger$     & BG            & 2011 Jan. 27      & 3.44  & 14.5  & 17.2\\
    BD+22~574   & CC1$^\dagger$     & BG            & 2009 Oct. 31      & 3.29  & 92.6  & 19.2\\
                & CC2$^\dagger$     & BG            & 2009 Oct. 31      & 8.50  & 50.0  & 17.4\\
    V1174~Tau   & CC1$^{\ddagger}$  & BG            & 2012 Sep. 12      & 6.47  & 63.6  & 18.0\\
                &                   &               & 2013 Oct. 16      & 6.51  & 63.2  & - \\
                &                   &               & 2013 Nov. 24      & 6.46  & 63.3  & -\\
                &                   &               & 2014 Oct. 7, 9    & 6.46  & 62.9  & -\\
                & CC2$^{\ddagger}$  & BG            & 2012 Sep.12       & 9.24  & 37.5  & 18.5\\
                &                   &               & 2013 Oct. 16      & 9.35  & 37.2  & -\\
                &                   &               & 2013 Nov. 24      & 9.26  & 37.3  & -\\
                &                   &               & 2014 Oct. 7, 9    & 9.28  & 36.9  & -\\
    V1054~Tau   & CC1$^{\ddagger}$  & BG            & 2012 Sep. 12      & 7.05  & 110.1 & 18.1\\
                &                   &               & 2013 Feb. 26      & $\ast$& $\ast$& $\ast$\\
                & CC2$^{\ddagger}$  & BG            & 2012 Sep. 12      & 7.33  & 75.9  & 16.0\\
                &                   &               & 2013 Feb. 26      & $\ast$& $\ast$& $\ast$\\
    BD+01~565   & CC1               & C (known)     & 2012 Nov. 5       & 10.98 & 14.3  & 7.28\\
    ome~Tau     & CC1               & BG            & 2014 Jan. 20      & 7.47  & 305.3 & 17.5\\
                &                   &               & 2014 Oct. 10      & 7.52  & 305.3 & - \\
                & CC2               & BG            & 2014 Jan. 20      & 9.47  & 51.8  & 16.2\\
                &                   &               & 2014 Oct. 10      & 9.50  & 51.5  & - \\
    GJ~212      & CC1               & BG (extended) & 2013 Jan. 2       & 6.51  & 195.2 & 18.6\\
                &                   &               & 2013 Oct. 17      & 6.08  & 197.4 & -\\
                & CC2               & BG            & 2013 Jan. 2       & 7.72  & 232.8 & 19.3\\
                &                   &               & 2013 Oct. 17      & 7.50  & 237.0 & -\\
    HIP~36624   & CC1               & BG            & 2013 Nov. 23      & 1.45  & 128.7 & 16.6\\
                &                   &               & 2015 Jan. 7       & 1.47  & 126.3 & -\\
                & CC2               & BG            & 2013 Nov. 23      & 9.67  & 316.6 & 20.8\\
                &                   &               & 2015 Jan. 7       & 9.67  & 317.0 & -\\
    CCDM~J08316+3458A & CC1         & C (known)     & 2013 Nov. 24      & 0.205 & 341.2 & 6.31\\
                &                   &               & 2015 Jan. 8       & 0.150 & 314.9 & -\\
                & CC2               & Artifact?     & 2013 Nov. 24      & 3.608 & 303.7 & 20.8\\
                &                   &               & 2015 Jan. 8       & $\star$ & $\star$ & $\star$ \\
                & CC3               & U             & 2013 Nov. 24      & 3.774 & 343.7 & 21.4\\
                &                   &               & 2015 Jan. 8       & 3.762 & 343.9 & -\\
    BD-13~2855  & CC1               & BG            & 2013 Jan. 1       & 8.374 & 332.2 & 17.4\\
                &                   &               & 2014 Jan. 20      & 8.495 & 332.5 & -\\
                &                   &               & 2014 Apr. 23      & 8.506 & 332.7 & -\\
                & CC2               & BG            & 2013 Jan. 1       & 11.135 & 331.6 & 13.2\\
                &                   &               & 2014 Jan. 20      & 11.250 & 331.8 & -\\
                &                   &               & 2014 Apr. 23      & 11.264 & 332.0 & -\\
                & CC3$^{\sharp}$    & -             & -                 & -     & -     & -\\
    BD+52~1638  & CC1               & U             & 2013 Feb. 26      & 1.194 & 185.1 & 8.62\\
    HR~4803     & CC1               & C (known)     & 2013 Jan. 2       & 2.571 & 198.1 & 8.53\\
    BD+22~2522  & CC1               & C (known)     & 2013 May 18       & 1.645 & 51.4  & 10.6\\
    GJ~516      & CC1               & C (known)     & 2014 Jun. 8       & 2.744 & 53.2  & 7.39\\
    HR~7451     & CC1               & BG            & 2012 Sep. 14      & 9.318 & 140.5 & 19.3\\
                &                   &               & 2014 Apr. 24      & 9.010 & 139.8 & -\\
    BD-00~4333  & CC1               & BG            & 2012 Nov. 5       & 7.098 & 16.9  & 19.0\\
                &                   &               & 2013 Oct. 16      & $\ast$& $\ast$& $\ast$\\
                &                   &               & 2014 Jun. 7       & 7.172 & 16.9  & -\\
                & CC2               & BG            & 2012 Nov. 5       & 8.959 & 14.8  & 17.5\\
                &                   &               & 2013 Oct. 16      & $\ast$& $\ast$& $\ast$\\
                &                   &               & 2014 Jun. 7       & 9.033 & 14.8  & -\\
\end{longtable}